\documentclass[aps,prc,twocolumn,showpacs,superscriptaddress]{revtex4-1}

\usepackage{amsmath,amssymb,amsfonts}
\usepackage[T1]{fontenc}
\usepackage{times} 
 \usepackage{graphicx}
\usepackage[margin=5pt, font=footnotesize, justification=raggedright, format=hang ]{caption}

\begin{document}


\title{Investigation of the Cross Section for dd Elastic Scattering \\and dd$\rightarrow$ n$^{3}$He Reactions at 160 MeV}



\author{I.~Ciepa\l}
\email[Electronic mail: ]{izabela.ciepal@ifj.edu.pl}
\affiliation{Institute of Nuclear Physics, PAS, PL-31342 Krak\'ow, Poland}
\author{J.~Kubo\'{s}}
\affiliation{Institute of Nuclear Physics, PAS, PL-31342 Krak\'ow, Poland}       
\author{K.~Bodek}
\affiliation{Institute of Physics, Jagiellonian University, PL-30348 Krak\'ow, Poland}
\author{N.~Kalantar-Nayestanaki}
\affiliation{KVI-CART, University of Groningen, NL-9747 AA Groningen, The Netherlands}
\author{G.~Khatri}
\affiliation{Department of Physics and Astronomy, Northwestern University, Evanston, IL 60208, USA}
\author{St.~Kistryn}
\affiliation{Institute of Physics, Jagiellonian University, PL-30348 Krak\'ow, Poland}
\author{B.~K\l os}
\affiliation{Institute of Physics, University of Silesia, PL-41500 Chorz\'{o}w, Poland}
\author{A.~Kozela}
\affiliation{Institute of Nuclear Physics, PAS, PL-31342 Krak\'ow, Poland} 
\author{P.~Kulessa}
\affiliation{Institute of Nuclear Physics, PAS, PL-31342 Krak\'ow, Poland}               
\author{A.~Magiera}
\affiliation{Institute of Physics, Jagiellonian University, PL-30348 Krak\'ow, Poland}
\author{J.~Messchendorp}
\affiliation{KVI-CART, University of Groningen, NL-9747 AA Groningen, The Netherlands}
\author{I.~Mazumdar}
\affiliation{Tata Institute of Fundamental Research, Mumbai 400 005, India}
\author{W.~Parol}
\affiliation{Institute of Nuclear Physics, PAS, PL-31342 Krak\'ow, Poland}
\author{D. Rozp\k{e}dzik}
\affiliation{Institute of Physics, Jagiellonian University, PL-30348 Krak\'ow, Poland}
\author{A.~Rusnok}
\affiliation{Institute of Physics, University of Silesia, PL-41500 Chorz\'{o}w, Poland}
\author{I.~Skwira-Chalot}
\affiliation{Faculty of Physics University of Warsaw, PL-02093 Warsaw, Poland}                 
\author{E.~Stephan}
\affiliation{Institute of Physics, University of Silesia, PL-41500 Chorz\'{o}w, Poland}
\author{A. Wilczek}
\affiliation{Institute of Physics, University of Silesia, PL-41500 Chorz\'{o}w, Poland}  
\author{B.~W\l{}och}
\affiliation{Institute of Nuclear Physics, PAS, PL-31342 Krak\'ow, Poland}    
\author{A. Wro\'{n}ska}
\affiliation{Institute of Physics, Jagiellonian University, PL-30348 Krak\'ow, Poland}
\author{J.~Zejma}
\affiliation{Institute of Physics, Jagiellonian University, PL-30348 Krak\'ow, Poland}

\date{\today}

\begin{abstract}
Differential cross sections of $^{2}$H({\em d},{\em d}){\em d} elastic scattering and proton transfer $^{2}$H({\em d},$^{3}$He){\em n} reactions 
at 160~MeV beam energy have been obtained.
 They have been normalized relative to the existing cross-section data for the $^{2}$H({\em d},{\em d}){\em d} elastic scattering at 180 and 130~MeV, 
 benefitting from the negligible
energy dependence of this observable at certain range of the four momentum transfer.
The experiment was performed at KVI in Groningen, the Netherlands using the BINA detector.  
The elastic scattering data are compared to theoretical predictions based on the lowest-order term in the Neumann series expansion 
for four-nucleon transition operators. The calculations underpredict the data. 
The data presented in this paper can be used to validate the future theoretical findings.
 
\end{abstract}


\maketitle

\section{Introduction}
\label{intro}
Nuclear forces have been well developed in the past two decades. 
According to our present knowledge, the nuclear force is due 
to residual strong interactions between the colorless hadrons - similar
to the van der Waals force between neutral atoms. 
The first idea for describing the nuclear force was developed by Yukawa, who
suggested that the force is generated via exchange of massive particles (namely pions which were later discovered) 
between the nucleons \cite{Yuk:35} - in analogy to 
the electromagnetic interaction where the exchange of a massless photon creates the force of infinite range. 
On the basis of the theory, models of nucleon-nucleon (NN) forces were created.
Nowadays, they are still in use in the sophisticated form of so-called realistic potentials, since the theory of strong interactions, QCD,
cannot yet be solved in the nonperturbative regime in the energy domain where stable nuclei exist.  
These potentials can include $\Delta$-isobar degrees of freedom in the coupled-channels formalism \cite{Del:08}. 
There is also a more fundamental way of describing nuclear forces within 
Chiral Perturbation Theory (ChPT) \cite{Ent:03,Ent:17} - effective field theory based on the symmetries of QCD. 
All these approaches constitute a rich
theoretical basis for description of the NN interaction. To investigate the nature of the nuclear 
interactions, few-nucleon systems were chosen as a basic experimental laboratory. 
As a consequence of internal structure of the nucleon, the so-called three nucleon force (3NF) 
plays an important role in three-nucleon (3N) systems. Exact calculations for such systems are performed with various NN potentials and 3NF
models and are directly compared to experimental data, providing information on the quality of the interaction. 
In general, the bulk of {\em p}-{\em d} scattering data is well described by the predictions. Remaining discrepancies 
are observed at low energies (e.g. the A$_{y}$ anomaly or the symmetric star anomaly \cite{Sag:10}) and at intermediate  energies \cite{Kal:12}. 
The latter ones might be partially caused by still non-relativistic
treatment of 2N and 3N scattering problem.\\
\indent In heavier systems composed of four nucleons (4N), larger sensitivity to the 3NF effects is expected.
This makes the experimental studies attractive, however the theoretical treatment of 4N scattering at intermediate energies 
(well above the breakup threshold) is much more complicated and challenging than for 3N systems \cite{Fonseca:17}.             
Such 4N ensemble reveals the complexity of heavier systems \cite{Deltuva:07a}, e.g. a variety of entrance and exit channels, various 
total isospin states etc.. 
The neutron-$^{3}$H~({\em n}-$^{3}$H) and proton-$^{3}$He ({\em p}-$^{3}$He) scattering are dominated by the 
total isospin {\em T}=1 states while {\em d}-{\em d} scattering has only {\em T}=0 component; 
the reactions {\em n}-$^{3}$He and {\em p}-$^{3}$H involve both {\em T}=0 and {\em T}=1 and are coupled to {\em d}-{\em d} in {\em T}=0, also a small 
admixture of {\em T}=2 states is present due to 
the charge dependence of the interactions.
The Coulomb force results in not only the repulsion
but also the splitting of {\em n}-$^{3}$He and {\em p}-$^{3}$H thresholds.
Another important aspect of calculations involving 4N is the possibility to probe states in
the continuum associated with specific resonances. Such states may possess higher angular momentum than corresponding bound
states. All these features make the 4N scattering problem a perfect theoretical and experimental laboratory 
to test various nuclear potentials. It is also important to perform \textit{ab initio} calculations 
of the 4N systems after the long and extensive work on the 3N scattering.
The calculations involving 4N are mainly developed by three groups: Pisa \cite{Viviani:11, Kiev:08} and
Grenoble-Strasbourg \cite{Lazauskas:04, Lazauskas:09}, working in the coordinate space representation, and Lisbon-Vilnius group, 
using the momentum space Alt, Grassberger, Sandhas (AGS) equations for transition operators \cite{Fonseca:17}.
All these methods include the Coulomb force but only the Lisbon-Vilnius group 
calculates observables for multi-channel reactions above the breakup threshold.\\
\indent Theoretical calculations of the bound states show very weak influence of the 4N force so 
its contribution can be neglected \cite{Nogga:02, Del:08}. 
Nevertheless, one should verify this claim in the case of nuclear reactions. 
Recent years have brought tremendous progress in precise calculations 
of the cross section and polarization observables. The very first 4N scattering results with the realistic 2N forces 
were obtained for single channel {\em n}-$\!^{3}\mathrm{H}$, {\em p}-$\!^{3}\mathrm{He}$ \cite{Viviani:11, Kiev:08, Lazauskas:04} 
and {\em p}-$\!^{3}\mathrm{H}$ \cite{Lazauskas:09} reactions below inelastic threshold.
Then, the observables were calculated at energies below the three-body breakup threshold, for {\em n}-$\!^{3}\mathrm{H}$ \cite{Deltuva:07a}, 
{\em p}-$\!^{3}\mathrm{He}$ \cite{Deltuva:07b},
 {\em n}-$\!^{3}\mathrm{He}$, {\em p}-$\!^{3}\mathrm{H}$ and {\em d}-{\em d} \cite{Deltuva:07c}. 
So far, the rigorous predictions have been limited to a domain of the lowest energies. 
The {\em p}-$\!^{3}\mathrm{He}$ is the simplest, since it involves three protons and one neutron, 
therefore only elastic and breakup channels exist. On the other hand, 
the most serious complication for {\em p}-~$\!^{3}\mathrm{He}$ is due to 
the Coulomb interaction between protons, which is treated using the method of screening and renormalization \cite{Deltuva:13}.
\\
\indent Recently, the calculations were extended to energies above the four-cluster breakup threshold, up to 35~MeV. 
The system dynamics is modeled with various potentials enabling studies of the sensitivity to the dynamics.
The following potentials were utilized  in the calculations:
CD Bonn \cite{Mach:01} and Argonne V18 (AV18) \cite{Wir:95} potentials, 
INOY04 (the inside-nonlocal outside-Yukawa) potential by Doleschall \cite{Dol:04}, potential  derived from ChPT at N3LO \cite{Ent:03} and
the two-baryon coupled-channel potential CD Bonn+$\Delta$ \cite{Deltuva:03}. The last potential yields effective 
three- and four-nucleon forces \cite{Del:08}. 
The sensitivity to the force model is rather small at the energy range studied. 
The predictions have been made for observables in {\em p}-$\!^{3}\mathrm{He}$ \cite{Deltuva:13}, {\em n}-$\!^{3}\mathrm{He}$ \cite{Deltuva:14a} 
elastic scattering and transfer reactions. The 3N and 4N dynamical components  were 
modeled separately via the explicit treatment of a single $\Delta$-isobar for {\em p}-$\!^{3}\mathrm{He}$ elastic scattering at 30~MeV.
The observables for {\em p}-$\!^{3}\mathrm{He}$ elastic scattering have also been calculated at proton beam energy of 70~MeV \cite{Deltuva:15}. 
Recent progress in calculations for {\em d}-{\em d} systems is  presented in \cite{Deltuva:15a, Deltuva:15b, Deltuva:16, Deltuva:17}
and in the review article~\cite{Fonseca:17}.\\
\indent  Calculations for the {\em d}-{\em d} system at higher energies are currently feasible and  
were performed in the so-called single-scattering approximation (SSA) for the three-cluster breakup 
and elastic scattering  \cite{Deltuva:16}. In this approximation instead of solving full AGS equations \cite{Gras:67}
the 4N operators are expanded in Neumann series in terms of 3N transition operators
and only the first-order contribution is retained.
This simplification is expected to be reasonable
only at higher energies. Similar approximation was already
used to calculate {\em d}-{\em d} elastic scattering at 230 MeV \cite{Mich:07}. In the calculations three different
2N potentials were used: AV18 \cite{Wir:95}, CD Bonn \cite{Mach:01} and CD Bonn+$\Delta$ \cite{Deltuva:03}.
To demonstrate the reliability of the SSA 
calculations for the 4N system, the same kind of approximation was applied to the {\em p}-{\em d} breakup \cite{Deltuva:16}.
The exact calculations for the 3N breakup were compared to ones obtained within SSA. 
The total {\em p}-{\em d} breakup cross section calculated in an exact way is lower than the one obtained 
in SSA by 30\% at 95 MeV and by 20\% at 200 MeV.\\
\indent The recent SSA calculations for the elastic scattering \cite{Deltuva:16} 
use more refined NN potentials \cite{Mach:01,Wir:95,Deltuva:03} 
and take into account more partial waves than in the 
previous calculations \cite{Mich:07},
therefore the 4N results are well converged.
Moreover, the interaction part
was improved by adding external Coulomb correction to the elastic scattering amplitude.
As the authors of  Ref.~\cite{Deltuva:16} conclude, the recent theoretical predictions should give 
correct orders of magnitude for total and differential cross sections for {\em d}-{\em d} and {\em p}-{\em d} breakup (near quasi-free region) 
and elastic scattering. \\
\indent The world database on 4N scattering at medium energies is still very poor. 
The data are often measured in very narrow angular ranges.
Since {\em n}-$\!^{3}\mathrm{He}$ experiments are difficult, the {\em p}-$\!^{3}\mathrm{He}$ 
and  {\em d}-{\em d} reactions are more often used to study 4N system. \\
\indent With the two deuterons in the initial state, in addition to the simple elastic scattering process, several 
reactions with a pure hadronic signature can occur: 
\begin{enumerate}
 \item neutron-transfer: {\em d}+{\em d} $\rightarrow$ {\em p}+$^{3}$H,
 \item proton-transfer: {\em d}+{\em d} $\rightarrow$ {\em n}+$^{3}$He,
 \item three-body breakup: {\em d}+{\em d} $\rightarrow$ {\em d}+{\em p}+{\em n},
 \item four-body breakup: {\em d}+{\em d} $\rightarrow$ {\em p}+{\em p}+{\em n}+{\em n}.
\end{enumerate}
 For these reactions only few data sets exist at intermediate energies (50-230 MeV) 
 \cite{Alde:78,Bizard:80,Mich:07,Bail:09,Ram:09,Ram:11,Khatri:15}. \\
 \indent The $^{2}$H({\em d},$^{3}$He) reaction implies high momentum transfer. 
In such reactions, energy, momentum, and angular momentum are exchanged by the projectile and target nucleus, 
as in the case of inelastic scattering. However, in the transfer reaction there is also transfer of mass. 
Reactions of this type have been a very important tool for the study of nuclear structure and helped to validate the nuclear
shell model by identifying the single-particle states.\\
\indent In this article, measurements of the differential cross section at 160 MeV beam energy for the two channels of 
{\em d}-{\em d} collisions: elastic scattering and proton transfer will be presented. 
\section{Detection System and Experiment}
\label{exp}
\begin{figure}[!h]
 \includegraphics[width=0.5\textwidth]{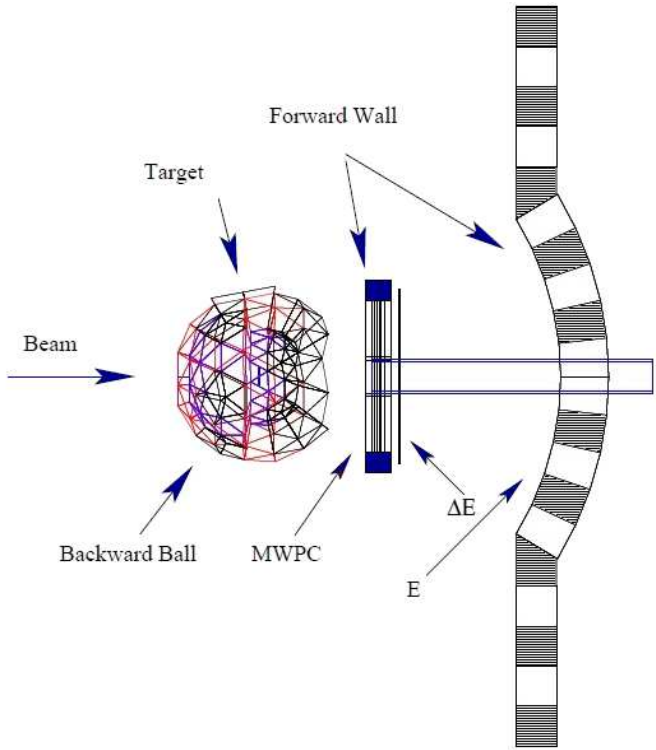}
 \caption{A schematic view of the BINA detector.}
 \label{fig1}
 \end{figure}

The experiment was carried out at Kernfysisch Versneller Instituut
(KVI) in Groningen, the Netherlands. The deuteron beam was provided by the superconducting
cyclotron AGOR (Accelerator Groningen ORsay) at kinetic energy of 160
MeV and was impinged on a liquid deuterium target. The nominal thickness of the target cell was 6.0 mm.
In addition, the thickness was increased by about 0.6 mm due to bulging of the
cell which leads to the effective target surface density of 107.2 $\pm$ 3.3 mg/cm$^{2}$ \cite{Kal:98}.
Low beam currents (about 5 pA) were used in order to keep the level of accidental
coincidences as low as possible. The reaction products were detected using BINA - Big Instrument
for Nuclear Polarization Analysis \cite{Khatri:15,Mar:08,Stephan:13}, which inherits a number of its features
predecessor, the Small-Angle Large-Acceptance Detector (SALAD) \cite{Kal:00}. The BINA
detector was designed to study few-body scattering reactions at medium energies.
It allows to register coincidences of two-charged particles in nearly 4$\pi$ solid angle, making it possible
to study breakup and elastic scattering reactions. The detector is
divided into two main parts, the forward Wall and the backward Ball. A schematic view of the detection system is presented in Fig.~\ref{fig1}.

\subsection{Forward Wall}
The forward Wall is composed of a three-plane multi-wire proportional chamber (MWPC)
and an array of an almost-squre-shapred array $\Delta E$-{\em E}  telescopes formed by two crossed layers of scintillator hodoscopes (vertically placed
thin transmission-$\Delta E$ strips and horizontally placed thick stopping-{\em E} bars). The forward
Wall covers polar angles $\theta$ in the range of 10$^\circ$-35$^\circ$ with the full range of azimuthal angles $\varphi$. 
MWPC is used to determine the position of the passing particle. The energy detectors $\Delta E$ and {\em E} are
used for measuring the energies of the charged reaction products and particle identification.  
The accuracy of the angle reconstruction is 0.3$^{\circ}$ for $\theta$ and between 0.6$^{\circ}$ and 3$^{\circ}$ for $\varphi$.
The energy resolution is about 2\%.
MWPC and the hodoscopes, have a central hole to allow for the passage of
beam particles to the beam dump. 
\subsection{Backward Ball}
\begin{figure}[!h]
 \includegraphics[width=0.25\textwidth]{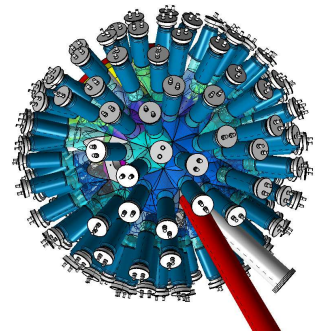}
 \caption{Ball part of the BINA detector.}
 \label{fig2}
 \end{figure}

The backward part of the detector is ball-shaped and consists of 149 scintillators, see Fig.~\ref{fig2}. 
The Ball plays both roles of a particle detector and a scattering chamber. 
It registers charged particles scattered at polar angles in the range of 40$^\circ$ to 165$^\circ$
with almost full azimuthal angle coverage. The shape and the construction of the inner surface of the Ball
can be compared to the surface of a soccer ball and is composed of 20 identical hexagon
and 12 identical pentagon structures. These polygons are further divided into isosceles triangles, thus spliting
the pentagon into five triangles and the hexagon into six triangles. Each triangle represents here a single Ball
detector. Ball elements placed at backward angles (larger than 90$^\circ$) 
are 3 cm thick, while those placed at more forward angles are 9 cm thick, allowing to stop protons up 
to 64 and 120 MeV, respectively.
The angular resolution of the Ball is set by its granularity thus it is worse than the one of the Wall. 
Depending on its orientation, one single element covers an angular
range up to $\pm$10$^\circ$, in both $\theta$ and $\varphi$ directions. 
Moreover, as the walls between single elements are not completely light tight, 
the scintillation light escapes to neighboring
elements. Therefore, in order to fully exploit the information about the energy deposited and the position of detected particle,  
it is necessary to consider a cluster as a "basic element" instead of one single scintillator
in the track reconstruction procedure. \\
\section{Data Analysis}
\label{data}
\begin{figure}[!h]
 \includegraphics[width=0.4\textwidth]{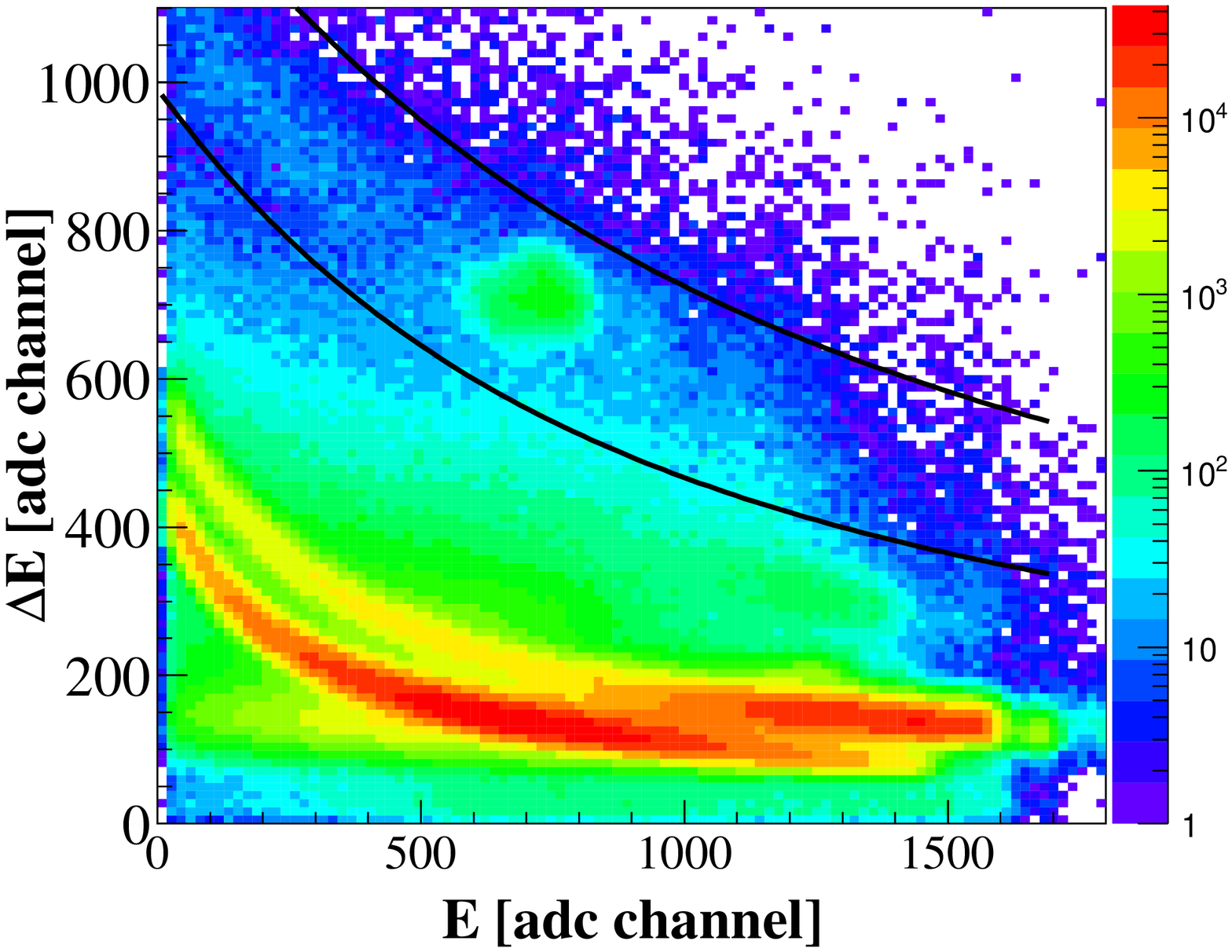}
 \includegraphics[width=0.4\textwidth]{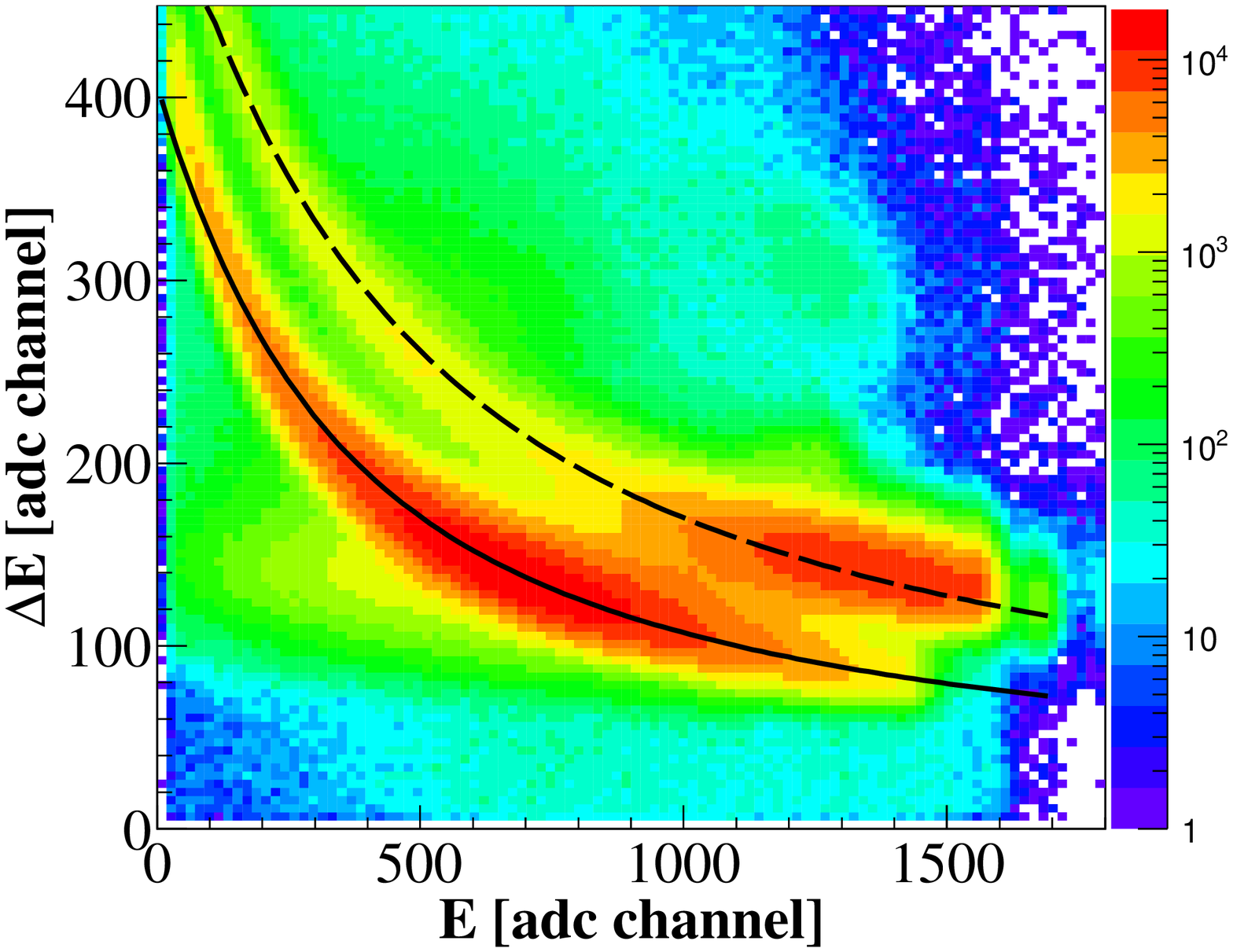}
 \caption{(Color online) {\em Upper panel:} Sample of $\Delta E$-{\em E} distribution with three sigma limits
 on $^{3}$He-ions. {\em Bottom panel:} The same as in the upper panel but zoomed in {\em dE} scale to show the fit 
 quality for protons and deuterons presented by solid and dashed lines, respectively.}
 \label{fig3}
 \end{figure}

\begin{figure}[!h]
 \includegraphics[width=0.4\textwidth]{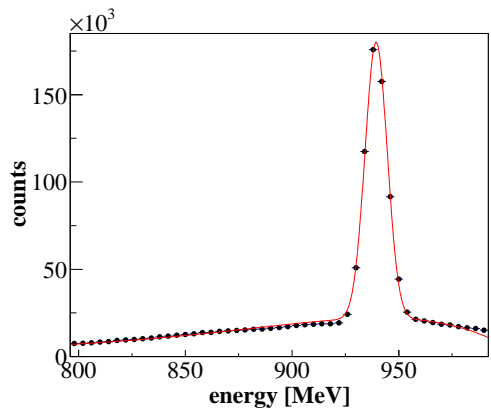}
 \caption{(Color online) Missing mass spectrum of the $d+d\rightarrow^{3}$He+X reaction at 160 MeV.
 A prominent peak corresponds to the mass of neutron.
 A signal (Gaussian) + background function (an 8$^{\rm th}$ order polynomial) was fitted to the
 distribution resulting in mean Gaussian of 939.4$\pm$0.1 MeV and $\sigma$=5.7$\pm$0.1 MeV.}
 \label{fig3b}
 \end{figure}

\begin{figure}[!hb]
  \includegraphics[width=0.4\textwidth]{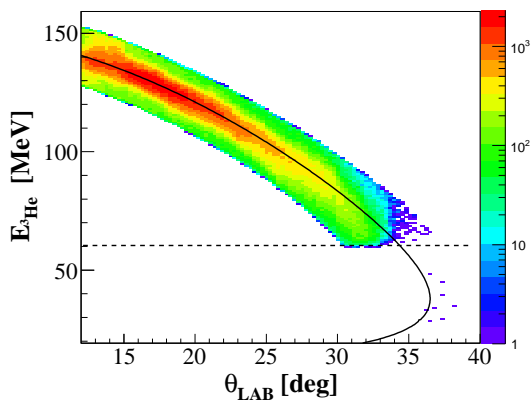}
 \caption{(Color online) Distribution of the reconstructed energy versus the polar angle for $^{3}$He-ions. 
 The solid black line represents the calculated kinematics. The dashed line indicates the energy threshold.}
 \label{fig4}
 \end{figure}

\begin{figure}[!h]
 \includegraphics[width=0.4\textwidth]{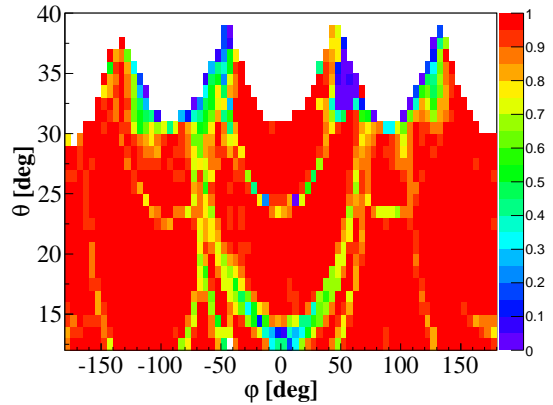}
 \caption{(Color online) The efficiency map of MWPC for the $^{3}$He-ions. The ellipse-like structures illustrate the ``dead'' 
 and malfunctioning wires.}
 \label{fig5}
 \end{figure}

\begin{figure}[!h]
 \includegraphics[width=0.35\textwidth]{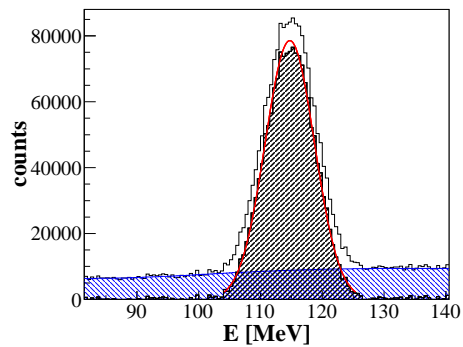}
 \caption{(Color online) A sample energy distribution of the $^{3}$He-ions at $\theta$=22$^{\circ}$ within a range of 1$^{\circ}$.
 The background is marked as the blue hatched area. The distribution obtained after the background subtraction 
 was fitted with a Gaussian function and integrated in the range of $-3\sigma$ and $+3\sigma$ from the peak position.}
 \label{fig6}
 \end{figure}

\begin{figure}[!h]
 \includegraphics[width=0.35\textwidth]{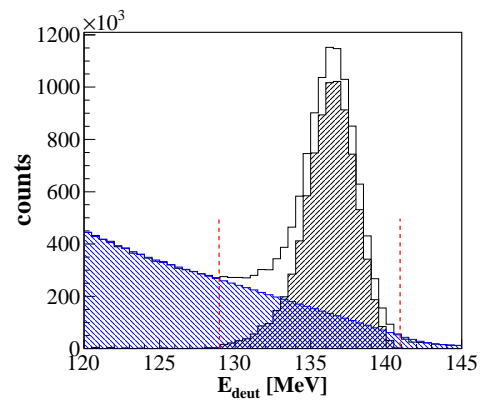}
 \caption{(Color online) A sample energy distribution of the deuterons registered at $\theta$=22$^{\circ}$ in a window of 2$^{\circ}$. 
 The background due to the breakup reaction is marked as the blue hatched area. The dashed lines 
 represent the integration region for the events.  
 }
 \label{fig7}
 \end{figure}

\begin{figure}[!h]
 \includegraphics[width=0.35\textwidth]{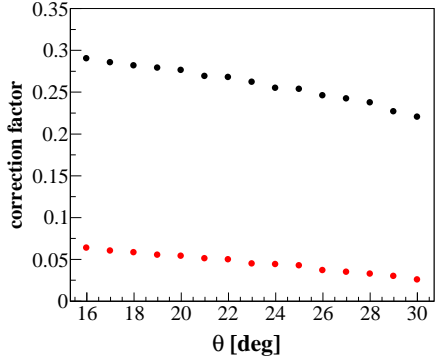}
 \caption{(Color online) Relative loss of events due to hadronic interaction of deuterons (black dots) and $^{3}$He-ions (red dots)
 in the plastic scintillator, presented as a function of the scattering angle. 
 }
 \label{fig8}
 \end{figure}

The particles of interest in the analysis presented here are elastically-scattered deuterons and $^{3}$He-ions 
from the proton-transfer reaction.
The events, which can be distinguished within the acceptance of the BINA detector are 
the Wall-Ball coincidences of the two deuterons from the elastic scattering, and the $^{3}$He 
single tracks (neutrons have not been reconstructed so far).  
The elastic scattering events have also been registered as single particles in the Wall (with minimum bias trigger).\\
\indent The very first step of the data analysis was event selection. 
The $\Delta E$-{\em E} method was used for particle identification (PID).
A banana-shaped cut for each individual $\Delta E$-{\em E} telescope 
is sufficient to separate protons and deuterons branches.
The cuts were defined wide enough to avoid significant losses of the particles. \\
\indent Alternatively, the so-called linearization method~\cite{Plos:75,Parol:14} was applied.
It relies on the introduction of a new variable:
\begin{eqnarray}
L=(dE+E)^\gamma-E^\gamma
\label{eq0}
\end{eqnarray}
where {\em dE} and {\em E} are energy losses in the thin and thick detectors, respectively.  
{\em L} is approximately independent of particle energy losses and the parameter $\gamma$ is specific for each 
virtual telescope (however, close to 1.76). $\gamma$ is adjusted to get a constant value for {\em L}, which leads to the best 
separation between protons and deuterons.
Then a Gaussian function has been fitted to the peaks in the distribution of {\em L} corresponding to each particle. 
Center values and widths obtained from the fit were subsequently used for particle selection (see Fig.~\ref{fig3}).
A given value of {\em L} transforms to  a bent line in $\Delta E$-{\em E} spectrum. 
The events within $\pm$2$\sigma$ in the {\em L} distribution were accepted for further analysis (see also Sec.~\ref{unc} where 
systematic uncertainties are discussed).
Although, since this method is based on protons and deuterons, much more abundant in PID spectra than $^{3}$He-ions, 
it has turned to be a very useful tool for consistent selection of $^{3}$He-ions registered in various telescopes. 
To check the correctness of the data selection procedure, the missing-mass spectrum of the $d+d\rightarrow^{3}$He+X reaction was     
drawn, see Fig.~\ref{fig3b}. \\
\indent After introducing PID into the analysis, the energy calibration was performed
for each type of particles. 
The proton energy calibration was based on elastically-scattered protons 
registered in dedicated measurements utilizing energy degraders of 
a few precisely controlled thicknesses and compared to GEANT4 simulations~\cite{Parol:14a,Kub:18}.  
Then, the position dependent information on average proton energy losses ($E_{\mathrm{dep}}^{\mathrm{p}}$) was used to obtain 
the proton energy at the reaction point ($E_{\mathrm{T}}^{\mathrm{p}}$). 
Due to different scintillation light output for protons and heavier particles (deuterons and $^{3}$He-ions),
additional corrections have been applied in order to obtain energy deposited in the thick {\em E}-scintillator ($E_{\mathrm{dep}}^{^{3}\mathrm{He}}$).
To calculate the energy of $^{3}$He/deuterons at the reaction point, the energy losses of $^{3}$He/deuterons on the
way from the target point to the {\em E}-detector were calculated based on the known relation $E_{\mathrm{T}}^{\mathrm{p}}$($E_{\mathrm{dep}}^{\mathrm{p}}$)
for protons \cite{Kub:18}.
Finally, the accuracy of the calibration
was tested by checking how well the 
kinematical relation between the energy and the polar angle of $^{3}$He-ions is reconstructed, see Fig. \ref{fig4}.
Equally good agreement was found in the case of the {\em d}-{\em d} elastic scattering kinematics.
  \indent To calculate the cross section, it is necessary to take into account the inefficiency of the detection system
and to correct accordingly the numbers of deuterons and $^{3}$He-ions registered at a given polar angle $\theta$. 
In the case of the BINA setup, the largest inefficiency is related to detection of particles in 
MWPC. Certain channels were malfunctioning or ceased to function at all (``dead'' wires). 
To compensate the experimental counting rates, the detector acceptance was divided into bins in azimuthal and 
polar angles and in this representation the position-dependent efficiency maps were constructed. The active part of MWPC contains three planes: 
with vertical and horizontal wires, and with wires inclined by 45$^{\circ}$. 
The efficiency map of each plane was obtained using the information from the remaining two others \cite{Parol:14}
and combining it with the information from the scintillator hodoscopes. 
The cumulative efficiency of MWPC for the $^{3}$He-ions is presented in Fig. \ref{fig5}. 
Similar maps were also calculated for deuterons and protons \cite{Parol:14}.
Much higher stopping power of $^{3}$He-ions results in about 5\% 
larger  average MWPC efficiency, as compared to deuterons.  
The difference between protons and deuterons is much less significant.   
In the case of the $\Delta E$ and {\em E} detectors the efficiency was established to close to 100\%.\\
\indent For the purpose of clean event selection, cuts were defined and imposed on the energy spectra.   
The events were sorted in  $\theta$ with the integration range of $\Delta\theta$=1$^{\circ}$
for the $^{3}$He-ions and $\Delta\theta$=2$^{\circ}$ for the deuterons.
Then the events were corrected for inefficiency and presented in the form of energy spectra. 
The background estimation and subtraction were performed 
with the use of the Statistics-sensitive Non-linear Iterative Peak-clipping (SNIP) algorithm \cite{Ryan:88}.  
A typical energy distribution for the $^{3}$He-ions is shown in Fig. \ref{fig6}. 
The main sources of the background are accidental coincidences, hadronic interactions inside the {\em E}-scintillators
and reactions induced by deuterons on passive material of the detector setup. 
The increase of the background  
on the high energy side can be attributed to the events leaking from neighboring triton line (see also Fig. \ref{fig3}). 
The Gaussian function was fitted to the final distributions and the events were integrated in the range 
corresponding to distances of -3$\sigma$ and +3$\sigma$ from the fitted peak.
 In the case of the elastically-scattered deuterons, background is due to the deuteron breakup and it was subtracted with the similar
algorithm as for the $^{3}$He-ions. A sample distribution is presented in Fig.~\ref{fig7}. In this 
case due to non-Gaussian character of the distributions, the events were integrated around the main peak
in the ranges defined separately for each $\theta$ angle. 
Finally, number of events counted at the given polar angles was corrected for losses due to the hadronic interactions inside 
the scintillator. The losses were calculated for the deuterons and $^{3}$He-ions with the use of the GEANT4 framework
and are presented in Fig.~\ref{fig8}. 

\subsection{Determination of Luminosity and dd Elastic Cross Section}
\label{norm}
\begin{figure}[!h]
 \includegraphics[width=0.5\textwidth]{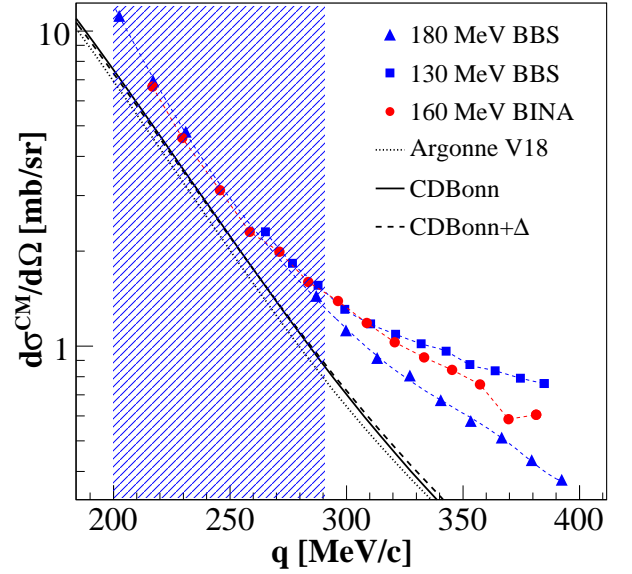}
 \caption{(Color online) The elastic scattering cross-section data measured at 160 MeV (red dots) are presented together 
 with the data measured in the  BBS experiment at 130 (blue squares) and 180 MeV 
 (blue triangles) \cite{Bail:09} (lines connecting points are used to guide the eye).
 Solid, dotted  and dashed black lines represent the calculations \cite{Deltuva:16} with the use of potentials specified in the legend. 
 The marked area refers to the region used in the normalization procedure.}
 \label{fig9}
 \end{figure}

\indent When presenting the {\em d}-{\em d} elastic scattering cross-section distribution as a
function of four-momentum transfer ({\em q}), 
one finds a scaling region where the distributions measured at different beam energies
overlap. Such an effect suggests simplicity of the reaction mechanism in which the
dynamical part of the scattering cross section predominantly depends on \textit{q} regardless
of the reaction energy. 
The four-momentum transfer \textit{q}, which is
square root of absolute value of the Mandelstam variable (\textit{t}),  is given as follows:
\begin{equation}
 q=\sqrt{\mid t-t_{max}\mid},\\ 
 \label{eq1a}
\end{equation}
\begin{equation}
 t-t_{max}=(\vec{p}_{d}^{~CM})^{2}\cdot(\cos\theta^{CM}_{d}-1),\\
 \label{eq1b}
\end{equation}
where $\vec{p}_{d}^{~CM}$ ($\theta^{CM}_{d}$) is the momentum 
(scattering angle) of the deuteron in the center-of-mass
frame, $t_{max}$ is the maximum value of {\em t} for cos$\theta^{CM}=0^{\circ}$ (see also Sec. \ref{cs}).
Benefiting from the existence of the scaling region, the data normalization was performed 
in the {\em q} range of 200-290 MeV/c, as shown in Fig. \ref{fig9}, 
by comparing the shape of elastic scattering rates measured in this experiment to 
the cross section obtained at 180 and 130 MeV \cite{Bail:09}. 
For the optimal scalling factor the maximum deviation was found to be 3\%.
As the result, the normalization factor was found to be $\kappa_{norm}$=(44.8$\pm$3.6(syst.))*10$^{6}$~[$\frac{1}{mb}$].
This parameter corresponds to the luminosity integrated over time of the experiment. It depends on the 
beam current, dead time of the data acquisition, the density and the thickness of the target.
The normalized data are presented in Fig.~\ref{fig9} (red dots) together with the recent 
calculations. Outside of the scaling region the cross section obtained in the present experiment 
fits well to the trend of previously measured data at two other energies.  
The theoretical predictions based on the Single-Scattering Approximation 
with AV18, CD Bonn and CD Bonn+$\Delta$ potentials
underestimate the data, 
as it was expected by the authors \cite{Deltuva:16,Deltuva:17a}.
\subsection{Differential Cross Section of d+d $\rightarrow$ n+$^{3}$He Transfer Reaction}
\label{cs}
\begin{figure}[!h]
  \includegraphics[width=0.5\textwidth]{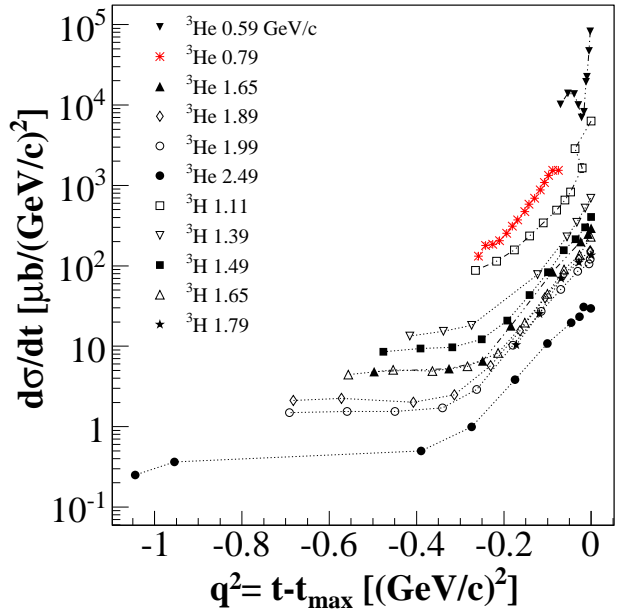}
  \caption{(Color online) Differential cross sections for transfer reactions in {\em d}-{\em d} systems 
    measured at various deuteron beam momenta, presented as a function of 
  $q^{2}$ \cite{Roy:69,Bizard:80}. The present data from the BINA experiment at 0.79 GeV/c 
 are shown as red stars. }
 \label{fig10}
 \end{figure}

\indent To compare the differential cross section of $d+d \rightarrow n+^{3}$He at 160 MeV with the existing database,
the data are presented as a function of square of four-momentum transfer $q^{2}=t-t_{max}$.
In this case the four-momentum transfer in the CM system is given by:
\begin{eqnarray}
 t &=&  (\vec{P}_{{}^{3}He} - \vec{P}_{beam})^{2}\nonumber \\
   &=&m_{d}^{2}+m_{{}^{3}He}^{2} -2\cdot E_{beam}^{CM}\cdot E_{{}^{3}He}^{CM}  \nonumber \\
   & &+2 \cdot \mid\vec{p}_{beam}^{~CM} \mid \cdot \mid\vec{p}_{{}^{3}He}^{~CM} \mid \cdot \cos\theta^{CM},
 \label{eq1}
\end{eqnarray}
where $\theta^{CM}$ is the $^{3}$He emission angle in the CM system.  
Therefore (see also Eqs. \ref{eq1a}, \ref{eq1b}):
\begin{equation}
t-t_{max}= 2\cdot \mid\vec{p}_{beam}^{~CM} \mid\cdot\mid\vec{p}_{{}^{3}He}^{~CM}\mid\cdot(\cos\theta^{CM}-1).
\label{eq2}
\end{equation}
 The $^{2}$H({\em d},$^{3}$He){\em n} cross section has been normalized with the use of the $\kappa_{norm}$ luminosity factor and the final distribution is presented
 in Fig.~\ref{fig10} together with the previous data from Refs. \cite{Roy:69,Bizard:80}. The present data follow the general trend
 of the distributions measured in the previous experiments.
  
\subsection{Experimental Uncertainties}
\label{unc}
\begin{figure}[!h]
 \includegraphics[width=0.45\textwidth]{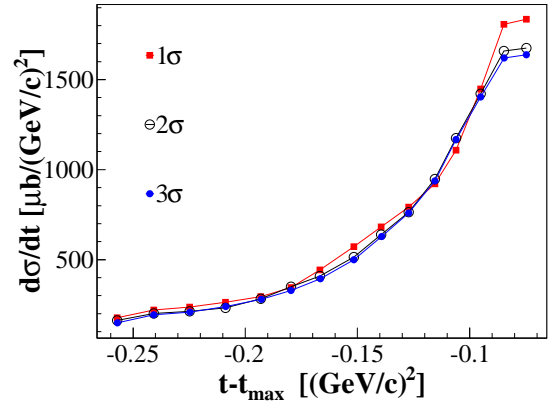}
 \caption{Differential cross section obtained for three different ranges of the linearization function {\em L}. 
 See text for details.}
 \label{fig11}
 \end{figure}

\begin{figure}[!h]
 \includegraphics[width=0.4\textwidth]{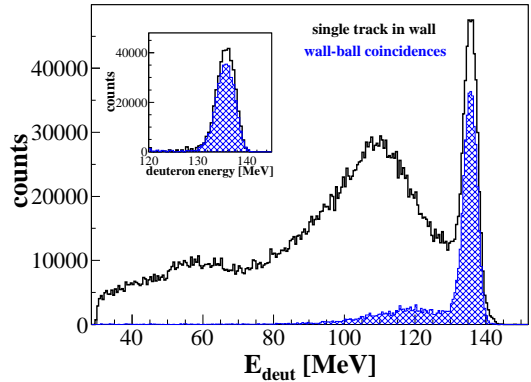}
 \caption{(Color online) A sample energy distribution of the elastically scattered deuterons 
 registered as the Wall-Ball coincidences (blue hatched distribution) for the Ball element 47 and an angular cover in the Wall of 
 ($\theta$=22$^{\circ}\pm0.5^{\circ}$, $\varphi$=$-7^{\circ}\pm5^{\circ}$). The distribution is compared to one obtained for the single tracks 
 in the Wall at the same polar angle.
Inset shows the same comparison after applying the background subtraction procedure as described in Sec.~\ref{data}.
 }
 \label{fig12}
 \end{figure}

The main sources of the systematic uncertainties which can affect the cross-section results
are related to the PID method, background subtraction and normalization procedure. To control systematic errors  
detailed studies of geometry of the setup and the detection efficiency were performed \cite{Khatri:15}.\\
\indent Protons and deuterons were identified via graphical cuts 
enclosing the branches/spots in the $\Delta E$-{\em E} spectra. 
The systematic uncertainty 
associated with this process was estimated by repeating the analysis based on modified
cuts and calculating the relative difference of the resulting
cross-section values. The typical uncertainty related to this effect, on the final
cross section, was found to be less than 3\%. \\
\indent In the case of the $^{3}$He-ions the relative size of the systematic
uncertainty due to PID has been found to be between 2 and 10\%, depending on the $q^{2}$ value. 
This has been estimated with the use of the cross-section distributions obtained for 
three different ranges of accepted $^{3}$He particles, 
defined as 1$\sigma$, 2$\sigma$ and 3$\sigma$  around the center of the corresponding peak in the linearization variable {\em L} (see Eq. \ref{eq0}),  
as presented in Fig. \ref{fig11}.
 The largest discrepancies appear at the highest $q^{2}$, 
 which correspond to the maximal polar angle of detected $^{3}$He-ions at which the 
 data are affected the most by the energy threshold, see Fig.~\ref{fig4}. \\
\indent In the case of the elastic-scattering process 
the background contribution due to the breakup reactions
is large and its subtraction procedure is a potential source of
significant systematic uncertainty. The associated errors were estimated 
based on the difference in the number of counts obtained in the two cases: 
for the single tracks in Wall (no correlation with a signal in the Ball required) and the Wall-Ball coincidences 
(with requirement of coplanarity and correct correlation of $\theta$ angles).
The Wall-Ball coincidences are much less affected by background, see Fig. \ref{fig12}, but depend on the Ball efficiency, 
which cannot be estimated in an independent way (without reference to the Wall). 
Therefore, the coincidences were defined for particularly well working Ball detectors which
were not affected by the energy threshold effects. 
For these events the deuteron energy distributions were
checked and the Gaussian function was fitted to energy distributions for the single events and Wall-Ball coincidences.
The relative difference between corresponding parameters of the Gaussian functions $\sigma_{single}$($\mu_{single}$) and $\sigma_{Wall-Ball}$($\mu_{Wall-Ball}$) is less than 10\%.  
This proves that the background is subtracted in a proper way and the systematic uncertainty was estimated to be around 4\%
(the SNIP algorithm~\cite{Ryan:88}). 
In the case of the $^{3}$He-ions, the background was estimated with the SNIP algorithm \cite{Ryan:88}
and the systematic error connected to this procedure was evaluated to be about 4\%.\\
\indent In the scaling procedure, we assume
that the cross section at different beam energies are equal (as shown in Fig.~\ref{fig9}),
showing almost energy independent behavior of the dynamics. The scaling is
not necessary exact, therefore 
experiments with the two beam energies close to each other were chosen for this purpose. 
The cross-section distribution for the elastic scattering was normalized to
the 180 MeV BBS data (the closest energy) and to the mixed data at 180 and 130 MeV
in the scaling region (see Fig.~\ref{fig9} and Sec.~\ref{norm}).
The systematic uncertainty was estimated as a deviation between results obtained in these two ways.
The maximum deviation reaches about 3\%. 
This value is further 
affected by the uncertainty in the BBS data, which is quoted to be 5\% in \cite{Bail:09}.\\
The other systematic effects are connected with the correction for the losses due to the hadronic interactions in the 
scintillator which was estimated to be less than 6\% and the losses due to the so-called crossover events \cite{Kis:05}.
Such events occur when particles penetrate from
one stopping detector to the adjacent one and in this case events are lost due to 
distorted energy information. In this experiment due to not proper light tightness between
{\em E}-slabs uncontrolled light leakage increased the crossovers. The systematic error was calculated
from the difference between the scaling factors obtained in the two cases with and without
taking into account the crossovers and is about 6\%. \\
\indent Summary of the systematic uncertainties is presented in~Tab.~\ref{tab1}. The total
systematic uncertainty composed of systematic errors added in quadrature vary between 13\% and 16\%.

\vspace{-4mm}
 \begin{table}[!h]
  \caption{Sources of systematic effects and their influence (in \%) on the final results.}
 \label{tab1}
 \begin{tabular}{|c|c|c|}
 \hline
 \textbf{Source of uncertainty} & \multicolumn{2}{c|}{\textbf{Size of the effect}} \\
                       & \textbf{p/d}& \textbf{$^{3}$He}    \\\hline                       
 PID & 3\% &2-10\%\\\hline
 Background subtraction & 4\% &4\% \\\hline
 Normalization & 5\%(BBS)+3\%&5\%(BBS)+3\% \\\hline
 Reconstruction of angles & 1\% &1\%\\\hline
 Energy calibration & 1\% &1\%\\\hline
 Hadronic interactions & 6\% &6\%\\\hline
 Crossover events & 6\% &6\%\\\hline
 \textbf{TOTAL} & \textbf{13\%} &  \textbf{13-16\%} \\\hline
  \end{tabular}
 \end{table}

\section{Conclusions and Outlook }
The differential cross-section distributions for the {\em d}-{\em d} elastic scattering (relative) and the $dd\rightarrow n^{3}$He 
proton transfer reaction have been obtained for deuteron-deuteron collisions at 160 MeV. \\
\indent The elastic-scattering data have been normalized 
to the earlier measurements in the overlapping range of momentum transfer.  
Outside of this range the data measured fit well to the trend observed at different energies.  
The cross section for the elastic scattering was compared
to the calculations based on the SSA \cite{Deltuva:16}. Its
validity is expected to improve with increasing
energy, but 160 MeV appears to be not high
enough. As expected, the calculations underpredict the data, 
however they provide correct order
of magnitude for the cross section.\\
\indent In the case of the proton transfer reaction, no calculations exist. However, the data presented can be used  
to validate the future theoretical findings. They supplement the existing database 
in the poorly known region of intermediate energies (beam momentum below 1 GeV/c). 
The  $q^{2}$ distribution follows the trend observed at higher energies obtained in proton and neutron transfer reactions.  \\


\begin{acknowledgments}
This work was supported by the Polish National Science Center under
Grants No. 2012/05/E/ST2/02313 (2013-2016) and 2016/21/D/ST2/01173 (2017-2020),
and by the European Commission within the Seventh Framework Programme
through IA-ENSAR (contract No. RII3-CT-2010-262010).
\end{acknowledgments}

\end{document}